\documentclass{IEEEtran}
\usepackage{amsmath,amssymb,amsfonts}
\usepackage{algorithmic}
\usepackage{graphicx}
\usepackage{array}
\usepackage{textcomp}
\usepackage{xcolor}  
\usepackage{url}
\usepackage[ruled,vlined,linesnumbered]{algorithm2e}
\usepackage{tikz}
\usepackage{booktabs}
\usepackage{graphicx}
\ifCLASSOPTIONcompsoc
\usepackage[caption=false, font=normalsize, labelfont=sf, textfont=sf]{subfig}
\usepackage{subcaption}
\else
\usepackage[caption=false, font=footnotesize]{subfig}
\fi
\usepackage{booktabs}
\usepackage{multirow}
\usepackage{cite}
\usetikzlibrary{arrows.meta,positioning,shapes.geometric}
\DeclareMathOperator*{\MAX}{MAX}
\usepackage[colorlinks=true, urlcolor=blue, linkcolor=blue,citecolor=blue]{hyperref}
\def\BibTeX{{\rm B\kern-.05em{\sc i\kern-.025em b}\kern-.08em
    T\kern-.1667em\lower.7ex\hbox{E}\kern-.125emX}}
\begin{document}
\title{PointNeRT: A Physics Aware Neural Ray Tracing Surrogate for Propagation Channel Modeling}
\author{	
	Zhuoyin~Li,
	Ruisi~He,~\IEEEmembership{Senior Member,~IEEE,} 
	Mi~Yang,~\IEEEmembership{Member,~IEEE,}
	Ziyi~Qi,
	Zhengyu~Zhang,~\IEEEmembership{Student~Member,~IEEE,}
	Jiahui Han, 
	Haoxiang Zhang,
	Bingcheng Liu
	\thanks{Zhuoyin Li, Ruisi He, Mi Yang, Ziyi Qi and Zhengyu~Zhang are with the School of Electronics and Information Engineering, Beijing Jiaotong University, Beijing 100044, China (e-mail: 25115062@bjtu.edu.cn; ruisi.he@bjtu.edu.cn; myang@bjtu.edu.cn; 22115006@bjtu.edu.cn; 98940447@bjtu.edu.cn).
		
	Jiahui Han and Haoxiang Zhang are with China Academy of Industrial
	Internet, Ministry of Industry and Information Technology, Beijing 100804, China (e-mail: hjh1760708@126.com; zhx61778294@126.com).
	
	Bingcheng Liu is with the Aerospace Information Research Institute, Chinese Academy of Sciences, Beijing 100094, China (e-mail: liubc@aircas.ac.cn).}
}

\maketitle

\begin{abstract}
Ray tracing (RT) has emerged as a key tool for propagation channel modeling and network planning. 
Conventional RT is based on electromagnetic (EM) wave theory and its application relies on detailed mesh-based environment representations and material properties. 
In realistic environments, limited environmental geometry and material uncertainties hinder its scalability to complex scenarios.
In this paper, we propose a novel physics aware neural RT surrogate named PointNeRT to address these limitations. 
The proposed model directly takes point clouds as environmental input, and efficiently reconstruct multipath without explicitly constructing mesh models or manually defining EM interaction rules.
PointNeRT adopts a hop-by-hop modeling strategy guided by physical interaction constraints. It supports sequential prediction of multipath propagation and power attenuation.
Numerical results and experiments demonstrate that the proposed method implicitly captures surface normal characteristics and EM material effects. It further achieves robust generalization in mobility scenarios and provides a physics-guided neural modeling of multipath propagation.

\end{abstract}

\begin{IEEEkeywords}
Radio propagation, ray tracing, channel modeling, point clouds.
\end{IEEEkeywords}

\section{Introduction}
\label{sec:introduction}
\IEEEPARstart{R}{ay} tracing (RT) is a fundamental technique used in propagation channel simulation to model electromagnetic (EM) wave propagation through explicit interactions within a given environment \cite{ref36, ref37}. Compared with widely used statistical channel models, RT provides physically consistent and spatially coherent propagation descriptions, which are essential for accurately capturing multipath behavior \cite{ref27}. Such properties are critical for many 6G research topics, such as integrated sensing and communications (ISAC) \cite{ref2}, \cite{ref41} radio-based localization \cite{ref3}, and wireless network planning and optimization \cite{ref4}. RT is also regarded as a key technology for wireless digital twin (DT) systems \cite{ref5}, \cite{ref42}.

Conventional RT is based on EM wave theory, and its applicability critically depends on mesh-based scene geometry and EM properties of surface materials \cite{ref1}.
In outdoor scenarios, geospatial data such as building models can be obtained from sources like Google Earth \cite{ref7} and OpenStreetMap. However, these datasets lack detailed structures of environments. 
In indoor scenarios, comparable geometric information is often unavailable \cite{ref6}.
As a result, existing RT tools primarily rely on manually constructed geometric models for detailed scene representation, which limits their generalization and maintainability \cite{ref38, ref39}. Moreover, manual modeling process itself is time-consuming, which further constrains practical applicability.
With recent advances in three-dimensional (3D) sensing technologies such as LiDAR, point clouds have emerged as an effective models of capturing accurate environmental representations. 
Point clouds directly encode scene geometry and offer advantages in acquisition speed and flexibility, making them well aligned with practical RT requirements \cite{ref28}.
Nevertheless, point cloud data are large scale and mesh reconstruction within conventional RT frameworks incurs significant computational overhead \cite{ref29, ref17, ref18}. 
To improve efficiency, subsequent studies accelerate this process by simplifying mesh resolution \cite{ref30}, \cite{ref40} or reducing ray launching density \cite{ref31}.
However, in complex scenes, such procedures still require substantial preprocessing and parameter tuning.
Consequently, achieving an effective balance between simulation accuracy and computational efficiency has become a critical challenge.

In this context, point-based RT on raw point clouds provides an alternative to mesh-based approaches.
It can avoid mesh reconstruction and retains high-fidelity geometric structures. There are essentially two different approaches to point-based RT in literature.
The first applies EM propagation theory directly to point clouds.
In \cite{ref32}, a single-lobe directive model that calculates EM field scattering from a small surface, is applied to point clouds. The same technique is used in \cite{ref24} to an outdoor-to-indoor scenario.
Although this approach reduces structure loss caused by mesh reconstruction to some extent, it still requires a trade-off between simulation accuracy and efficiency \cite{ref8}. To further improve computational performance, \cite{ref33} proposes a GPU-based radio wave propagation prediction method that exploits parallelism to speed up RT.

The second approach focuses on using neural networks (NNs) to learn how rays interact with points and thereby partially or fully replacing conventional RT computations.
A related framework is presented in \cite{ref16}, which applies traditional shooting and bouncing ray (SBR) simulations to target-level scenarios rather than full scene-level propagation modeling.
The authors of \cite{ref15} use three-dimensional Gaussian Splatting (3DGS) on point-based geometry to estimate received signal without explicitly modeling multipath propagation.
The idea has been extended to a physics-guided neural surrogate model in
\cite{ref34} for full channel impulse response (CIR) prediction.
The novelty with respect to prior work lies in the use of spherical harmonics to encode directional properties and Fourier Neural Operators, rather than treating NN as a black box, thereby enabling a physically informed neural framework.

The aim of this paper is to introduce a novel physics aware point cloud based neural RT surrogate (PointNeRT) that leverages a path-level propagation modeling to enable explicit representation of multipath trajectories and associated attenuation. The proposed approach takes point clouds as environmental input. Its output include both locations of interaction points and corresponding power attenuation. Instead of relying on entire scene point cloud, the network only need to extract local geometric features at each interaction point to characterize ray-environment interactions.
PointNeRT models multipath propagation as a sequence of hop-by-hop interactions, where each hop corresponds to a physically constrained propagation event. 
Benefiting from point-based environment representation and sequential propagation modeling framework, PointNeRT naturally scales to dynamic scenes and exhibits strong generalization capability.

The rest of this paper is organized as follows. Section \ref{sec:Physical Modeling and Problem Formulation} describes physical modeling and problem formulation. Section \ref{Point Cloud Based Neural Ray Tracing Architecture} details proposed PointNeRT framework. 
Section \ref{Network Implementation} contains training strategy and dataset preparation.
Section \ref{Experiments and Performance Evaluation} presents and discusses experimental results. Section \ref{Conclusion} concludes the paper.

\section{Physical Modeling and Problem Formulation}
\label{sec:Physical Modeling and Problem Formulation}
This section explains how CIR is obtained through RT simulation, which provides physical basis for hop-by-hop propagation modeling adopted in this work.

\subsection{Propagation Channels}
\label{Propagation Channels}
For a fixed link between a transmitter (Tx) and receiver (Rx), propagation delay and attenuation can be calculated.
The total field composed of $L$ propagation paths, each consisting of $B$ bounces, can be written as
\begin{equation}
	\label{equ0}
\mathcal{P}
=
\left\{
\left\{
a_{l,b},\, \tau_{l,b},\, \mathbf{I}_{l,b},\, (\theta_{l,b}, \phi_{l,b}),
\,(\theta'_{l,b}, \phi'_{l,b})
\right\}_{b=1}^{B}
\right\}_{l=1}^{L},
\end{equation}
where $a_l$ is received power of the $l$-th multipath component. $\mathbf{I}_{l,b} \in \mathbb{C}^{2 \times 2}$ is the polarimetric complex amplitude between two intersection point. $(\theta_l, \phi_l)$ and $(\theta'_l, \phi'_l)$ are the angles of departure and angles of arrival of the $l$-th path with delay $\tau_l$, respectively. 
The complex baseband-equivalent CIR at carrier frequency $f$ is calculated as a sum of propagation paths in (\ref{equ0}) with
\begin{align}
	h(\tau)
	&= \sum_{l=1}^{L} 
	\mathbf{G}'(\theta'_l, \phi'_l)\,
	\mathbf{T}_l \,
	\mathbf{G}(\theta_l, \phi_l)\,
	e^{-j 2 \pi f \tau_l}\,
	\delta(\tau - \tau_l), \label{eq:channel_model_full} \\
	&\triangleq \sum_{l=1}^{L} 
	a_l \, e^{-j 2 \pi f \tau_l}\,
	\delta(\tau - \tau_l). \label{eq:channel_model_compact}
\end{align}
where $\mathbf{G}(\theta_l, \phi_l)$ and $\mathbf{G}'(\theta'_l, \phi'_l)$ are transmit and receive antenna patterns, respectively. 
Matrix $\mathbf{T}_l : \mathbb{C}^2 \longmapsto \mathbb{C}^2$ are transfer function.
Using the Fourier transform, channel frequency responses is then given by
\begin{equation}
H(f) = \sum_{l=1}^{L} a_l\, e^{-j 2 \pi f \tau_l}.
\end{equation}

In point cloud based RT simulations, each interaction point is assumed to generate a scattered field, and total field is obtained by aggregating contributions from all points.

\subsection{Transfer Function}
\label{Transfer function}
The radio channel is estimated as a sum of $L$ signal paths between transmitter and receiver.
Each path is described as a hop-wise sequence of $B$ successive propagation and environment interaction events, including reflection, scattering, and diffraction.
For each interaction, propagation path from current interaction point $\mathbf{p}_b \in \mathbb{R}^3$ to next interaction point $\mathbf{p}_{b+1}$ involves both transmission loss and polarization transformation.
Let $\mathbf{E}^{\mathrm{out}}_b(\mathbf{p}_{b+1})$ denote outgoing electric field vector propagating from the $b$-th interaction point $\mathbf{p}_{b}$ toward $\mathbf{p}_{b+1}$.
Then, incident electric field at $\mathbf{p}_{b+1}$ is given by:
\begin{equation}
	\label{equ1}
	\mathbf{E}^{\mathrm{in}}_{b+1}(\mathbf{p}_{b+1})
	=
	\mathbf{D}_{b+1}\,
	\mathbf{E}^{\mathrm{out}}_b(\mathbf{p}_{b+1}),
\end{equation}
where $\mathbf{D}_{b+1}$ is a basis transformation matrix between adjacent interaction points.
Its explicit form is
\begin{equation}
	\mathbf{D}_{b+1}
	=
	\begin{bmatrix}
		(\hat{\mathbf{e}}^{\mathrm{in}}_{p,b+1})^{\mathrm{T}}\hat{\mathbf{e}}^{\mathrm{out}}_{p,b} &
		(\hat{\mathbf{e}}^{\mathrm{in}}_{p,b+1})^{\mathrm{T}}\hat{\mathbf{e}}^{\mathrm{out}}_{q,b} \\
		(\hat{\mathbf{e}}^{\mathrm{in}}_{q,b+1})^{\mathrm{T}}\hat{\mathbf{e}}^{\mathrm{out}}_{p,b} &
		(\hat{\mathbf{e}}^{\mathrm{in}}_{q,b+1})^{\mathrm{T}}\hat{\mathbf{e}}^{\mathrm{out}}_{q,b}
	\end{bmatrix},
\end{equation}
where $(\hat{\mathbf{e}}^{\mathrm{in}}_{p,b+1}, \hat{\mathbf{e}}^{\mathrm{in}}_{q,b+1})$ and
$(\hat{\mathbf{e}}^{\mathrm{out}}_{p,b}, \hat{\mathbf{e}}^{\mathrm{out}}_{q,b})$
are pairs of orthogonal unit vectors defining local polarization bases of incoming and outgoing fields at two adjacent interaction points, respectively.

At $\mathbf{p}_b$, outgoing electric fields $\mathbf{E}^{\mathrm{out}}_b(\mathbf{p}_{b+1})$ is given as
\begin{equation}
	\label{equ2}
	\mathbf{E}^{\mathrm{out}}_b(\mathbf{p}_{b+1})
	=
	\mathbf{F}_b\!\left(
	\mathbf{E}^{\mathrm{in}}_b(\mathbf{p}_b)
	\right),
\end{equation}
where $\mathbf{F}_b : \mathbb{C}^2 \rightarrow \mathbb{C}^2$ denotes polarimetric propagation operator associated with the $b$-th interaction, whose form depends on the type of scattering process.

Combining (\ref{equ1})--(\ref{equ2}), we obtain polarimetric complex amplitude at $\mathbf{p}_{b+1}$
\begin{equation}
	\label{equ3}
	\mathbf{I}_{b+1}
	\triangleq
	\mathbf{D}_{b+1}\mathbf{F}_b
	\in \mathbb{C}^{2\times2}.
\end{equation}
Then, the relationship between transmitted field $\mathbf{E}^{\mathrm{in}}_{B+1}(\mathbf{p}_{B+1})$ and received field $\mathbf{E}^{\mathrm{in}}_0(\mathbf{p}_0)$ has the following form
\begin{equation}
\mathbf{E}^{\mathrm{in}}_{B+1}(\mathbf{p}_{B+1})
=
\mathbf{D}_{B+1}\mathbf{F}_B
\left(
\cdots
\mathbf{D}_1\mathbf{F}_0
\mathbf{E}^{\mathrm{in}}_0(\mathbf{p}_0)
\right).
\end{equation}
According to (\ref{equ3}), above expression can be rewritten as
\begin{align}
	\mathbf{E}^{\mathrm{in}}_{B+1}(\mathbf{p}_{B+1})
	&=
	\mathbf{I}_{B+1}\mathbf{I}_{B}\cdots\mathbf{I}_1\,
	\mathbf{E}^{\mathrm{in}}_0(\mathbf{p}_0)
	\label{eq:field_representation_a} \\
	&=
	\mathbf{T}\!\left(\mathbf{G}(\theta,\phi)\right)\,
	e^{-j2\pi f \tau}.
	\label{eq:field_representation_b}
\end{align}
where $\tau = (\lambda f)^{-1} \sum_{b} d_b$ is total propagation delay, with
$d_b = \lVert \mathbf{p}_b - \mathbf{p}_{b-1} \rVert$.
The detailed derivation of above formulation can be 
found in \cite{ref19}.
In PointNeRT model, attenuation $ \mathbf{I}$ and propagation direction at each interaction point are explicitly predicted, from which the path parameters, such as $a_l$ and $(\theta_l, \phi_l)$, are obtained.
\subsection{Propagation Mechanisms}
\label{propagation mechanisms}
In our work, propagation mechanisms are divided into deterministic and non-deterministic interactions based on whether incident and outgoing directions of rays are strictly constrained by physical laws. This distinction allows model to better capture physical behavior of each mechanism using separate neural networks. Deterministic interactions are defined to include reflection, and non-deterministic interactions correspond to scattering and diffraction. Arbitrary multipath can be constructed by sequentially combining these interactions.
\subsubsection{Deterministic interactions} The power distribution of different propagation waves is constrained by the principle of energy conservation.
Let $R^2 \in [0,1]$ denote specularly reflected
fraction of reflected energy, and $S^2$ is scattering coefficient, defined as ratio of scattered power to incident power, where $S^2 = 1 - R^2$.

Consider a multipath component impinging on a planar surface at $\mathbf{p}_{b+1}$, with incident wave vector $\hat{\mathbf{k}}_i$ and surface normal vector $\hat{\mathbf{n}}$. 
We have
\begin{equation}
	\mathbf{I}_{b+1}
	=
	\mathbf{D}_{b+1}
	\begin{bmatrix}
		R r_{\perp} & 0 \\
		0 & R r_{\parallel}
	\end{bmatrix}
	A_r(\mathbf{p}_{b+1}, \mathbf{p}_{b})\,
	e^{-j \frac{2\pi}{\lambda} d_{b+1}},
\end{equation}
where $A_r(\cdot)$ is spreading factor related to the shape of propagation waves.
The Fresnel reflection coefficient $r$ depends on the material conductivity $\sigma$ and relative permittivity $\varepsilon_r$.
$\mathbf{D}_{b+1}$ can be found in \cite{ref20}. The direction of reflected waves $\hat{\mathbf{k}}_r$ is constrained by a fixed mapping $
\hat{\mathbf{k}}_r
=
\hat{\mathbf{k}}_i
-
2 \left( \hat{\mathbf{k}}_i^{\mathsf{T}} \hat{\mathbf{n}} \right) \hat{\mathbf{n}} 
$.
In reflection prediction model of PointNeRT, the network is designed to learn this deterministic geometric relationship, enabling accurate inference of outgoing direction under specular reflection constraints.

\subsubsection{Non-deterministic interactions}
For non-specular interactions, including diffuse scattering and diffraction, the polarimetric complex amplitude at scattering point can be computed according to \cite{ref21}:

\begin{equation}
	\begin{aligned}
		\mathbf{I}_{b+1}
		&=
		\mathbf{D}_{b+1}
		\frac{S\Gamma}{d_{b+1}}
		\int
		\sqrt{
			f_s(\hat{\mathbf{k}}_i,\hat{\mathbf{k}}_s,\hat{\mathbf{n}})
			\cos(\theta_i)
		}\,
		\mathrm{d}A \\
		&\quad \times
		\begin{bmatrix}
			\sqrt{1-K_x}\, e^{j\chi_1} \\
			\sqrt{K_x}\, e^{j\chi_2}
		\end{bmatrix}
		e^{-j2\pi d_{b+1}/\lambda}.
	\end{aligned}
\end{equation}
where $\Gamma^2$ is the fraction of incoming power, including both specular and diffuse components. $\chi_1, \chi_2 \in [0, 2\pi]$
represent independent random phase shifts, while $K_x \in [0, 1]$ is cross polarization discrimination coefficient. Differential area element $\mathrm{d}A$ corresponds to a small surface patch on reflecting object, and $f_s(\cdot)$ is scattering pattern.
Basis transformation matrix $\mathbf{D}_{b+1}$ ensures correct transfer of energy between two polarization components.

At diffraction points, the field transformation is modeled based on the uniform theory of diffraction \cite{ref22}. Following heuristic extension of diffraction model to finitely conducting wedges \cite{ref23}, polarimetric complex amplitude after diffraction can be given by
\begin{equation}
	\mathbf{I}_{b+1}
	=
	\mathbf{D}_{b+1}\,
	\mathbf{T}\,
	A_d(\mathbf{p}_{b+1}, \mathbf{p}_b)\,
	e^{-j2\pi d_{b+1}/\lambda},
\end{equation}
where $\mathbf{T}$ is diffraction matrix derived from diffraction coefficients and $A_d(\cdot)$ is diffraction spreading factor. By the law of edge diffraction, angles $\beta_0$ and $\beta_0'$ between edge and incident and diffracted rays, respectively, satisfy $\cos(\beta_0')=\cos(\beta_0)$. Consequently, diffracted rays are distributed on Keller’s cone, whose axis coincides with edge direction.

In non-deterministic interactions, propagation paths propagate following physically plausible patterns without a unique outgoing direction. The contribution of each multipath component to received signal is primarily determined by its attenuation. Consequently, directional power distributions are learned by the network rather than explicit directions.

\subsection{Path Tracing}
\label{Path tracing}

In this work, the dataset is generated by a conventional point cloud based RT simulator \cite{ref24}, which is calibrated with measurement data following the approach in \cite{ref26}.
EM material parameters $\sigma, \varepsilon_r, S, K_x$ are initialized using ``ITU Materials" \cite{ref25}.
To balance computational cost and physical fidelity, we impose explicit constraints on propagation paths. 
Specifically, we consider propagation paths with a maximum of three bounces and at most one diffuse reflection.
Only first-order diffraction paths are included, and refraction is neglected. 
In addition, arbitrary combinations of specular and diffuse reflections are allowed. The ray launching density is set to $10^{6}$.
For RT simulation model, diffuse reflection is modeled using Lambertian pattern and we do not apply random phase shifts to diffusely reflected paths, i.e., $\chi_1=\chi_2=0$. 
Path construction are performed using SBR method.

\section{PointNeRT: Point Cloud Based Neural Ray Tracing Architecture}
\label{Point Cloud Based Neural Ray Tracing Architecture}
This section introduces network architecture and design principles of PointNeRT, followed by loss functions.

\subsection{Main Framework}
\label{Main Framework}
\begin{figure*}[!htb]\centering
	\includegraphics[width=7in]{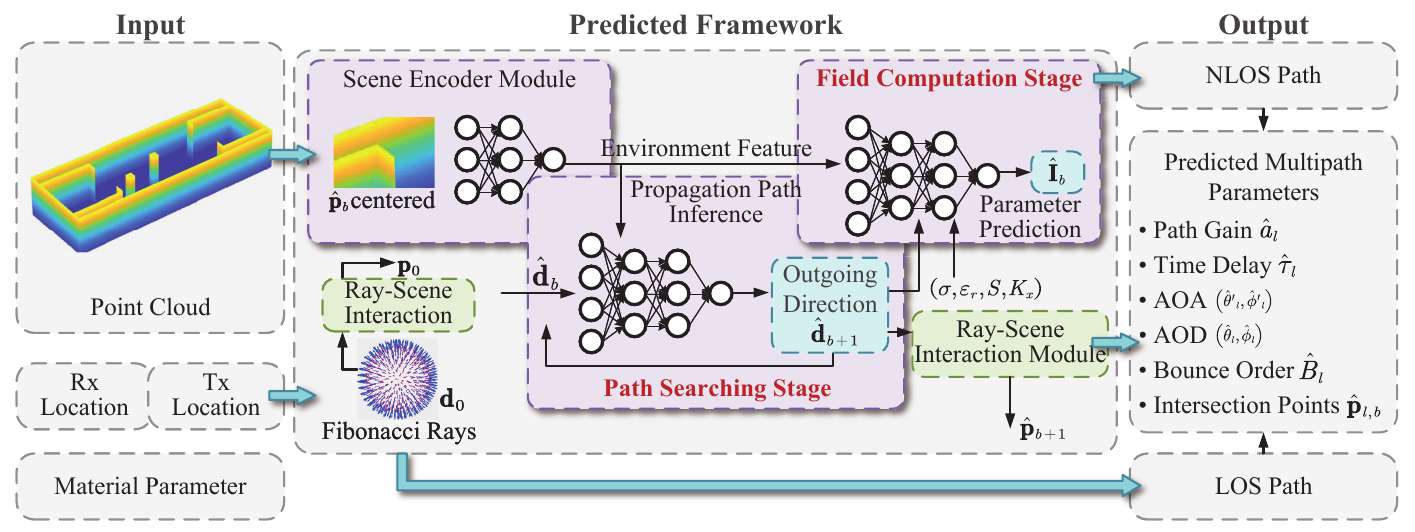}
	\caption{PointNeRT: Overall framework for predicting multipath parameters based on point clouds. Tx and Rx are employed to determine whether EM waves is received. Material parameters are used as inputs in field computation stage.}
	\label{figure1}
\end{figure*}

As illustrated in Fig.~\ref{figure1}, the proposed PointNeRT framework consists of three components: an input module, a physics aware prediction framework designed to model multipath propagation in a sequential manner, and an output module.

At input block, simulation scene is represented by three types of prior information: point cloud, Tx/Rx location, and EM material parameters. These inputs jointly characterize layout and physical properties of propagation environment.
Based on encoded scene representation, output part yields multipath parameters
$\{ a_l, \tau_l, (\phi_l^{r}, \theta_l^{r}), (\phi_l^{t}, \theta_l^{t}), B_l, \mathbf{p}_{l,b} \}$.

The prediction framework follows a physics consistent pipeline akin to a conventional RT simulator. It consists of two main stages: path searching and field computation. 
Before proceeding to these two main stages, the environment is first represented through geometric feature extraction.
Specifically, a scene encoder extracts environmental features from a local point cloud centered at interaction point $\mathbf{p}_b$. For first-order propagation, $\mathbf{p}_b$ is directly obtained from intersection between launched rays and point clouds. When $b>0$, $\mathbf{p}_b$ is determined according to predicted propagation direction $\mathbf{d}_{b+1} \in \mathbb{R}^3$.

With extracted local environmental features, path searching stage is then performed in an iterative manner.
At each subsequent order, it predicts outgoing direction $\mathbf{d}_{b+1}$ conditioned on current propagation direction $\mathbf{d}_b$ together with corresponding local environment representation. By iteratively applying this procedure, this stage progressively infers directional evolution of multi-bounce propagation paths.
Meanwhile, a ray-scene interaction module is employed at each order to test ray validity, determine whether a ray reaches Rx, and compute $\mathbf{p}_{b+1}$ for subsequent propagation.
Following path searching stage, field computation stage estimates polarimetric complex amplitude of each valid propagation path by jointly leveraging predicted direction, environmental features, and $(\sigma, \varepsilon_r, S, K_x)$. 

To account for different physical propagation mechanisms, deterministic and non-deterministic interactions are modeled using separate neural networks that share same architecture. By combining these two networks, propagation paths with mixed interaction attributes can be constructed.
In particular, the line of sight (LOS) component is obtained directly from geometric relationships, while non line of sight (NLOS) paths are inferred by NN.
The following part presents a detailed description of key modules in predicted framework.

\subsection{Predicted Network}
\label{Predicted Network}
\subsubsection{Scene Encoder Module}
In this module, we adopt simplified set abstraction (SA) levels inspired by PointNet++ to capture local geometric context at different spatial scales. 

At each level, a set of points
$
\mathcal{S} = \{\mathbf{x}_1, \mathbf{x}_2, \ldots, \mathbf{x}_N \mid \mathbf{x}_i \in \mathbb{R}^3\}
$
is processed and abstracted to produce a new point set with fewer elements. The set abstraction level is made of three key layers: A sampling layer, a grouping layer and MLP layer. 
At sampling layer, a farthest point sampling (FPS) operator $\mathcal{S}_{\mathrm{FPS}}(\cdot)$ is applied to select a subset of representative centroids $\mathcal{C}$, which is given by
\begin{equation}
	\mathcal{C}
	=
	\mathcal{S}_{\mathrm{FPS}}(\mathcal{S})
	=
	\{\mathbf{c}_j\}_{j=1}^{N'},
	\quad \mathbf{c}_j \in \mathcal{S}, \quad N' < N.
\end{equation}
Given centroid set $\mathcal{C}$ and original point set $\mathcal{S}$,  grouping layer constructs local neighborhoods $K$ within radius $r$ around each centroid $\mathbf{c}_j$. The grouping operator $\mathcal{G}(\cdot)$ is defined as
\begin{equation}
	\mathcal{G}(\mathbf{c}_j)
	=
	\left\{
	\mathbf{x}_i \in \mathcal{S}
	\;\middle|\;
	\lVert \mathbf{x}_i - \mathbf{c}_j \rVert \le r
	\right\},
	\quad j = 1, \ldots, N'.
\end{equation}
The output of grouping layer is a collection of local point sets
\begin{equation}
	\mathcal{G}(\mathcal{C})
	=
	\left\{
	\mathcal{G}(\mathbf{c}_j)
	\right\}_{j=1}^{N'},
\end{equation}
which corresponds to grouped point features of size $N' \times K \times (3 + C)$. $C$ is the point-wise feature dimension.
For the first SA module, $C = 0$.
These local groups are then processed by shared MLP layers $	\gamma_{\mathrm{cov2}}(\cdot)$ followed by symmetric aggregation to encode neighborhood information around each centroid. The resulting output feature matrix is given by
\begin{equation}
	\mathcal{F}
	=
	\left\{
	\MAX_{\mathbf{x}_i \in \mathcal{G}(\mathbf{c}_j)}
	\!\left\{
	\gamma_{\mathrm{cov2}}(\mathbf{x}_i - \mathbf{c}_j)
	\right\}
	\right\}_{j=1}^{N'}
	\in \mathbb{R}^{N' \times (3 + C')}.
\end{equation}

\begin{figure}[!t]\centering
	\includegraphics[width=3.5in]{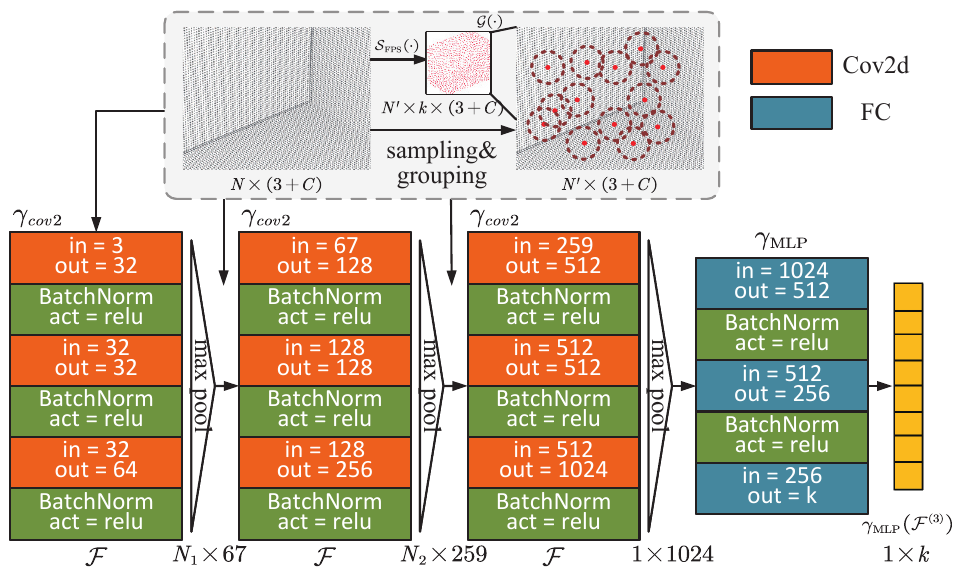}
	\caption{Structures of scene encoder module. }
	\label{figure2}
\end{figure}

After three hierarchical set abstraction levels, a final MLP is applied to obtain environment feature representation
$
	\gamma_{\mathrm{MLP}}\!\left(
	\mathcal{F}^{(3)}
	\right).
$
Further details of network structure are illustrated in Fig. \ref{figure2}.

\subsubsection{Propagation Searching Stage}

\begin{figure}[!htb]\centering
	\includegraphics[width=3.5in]{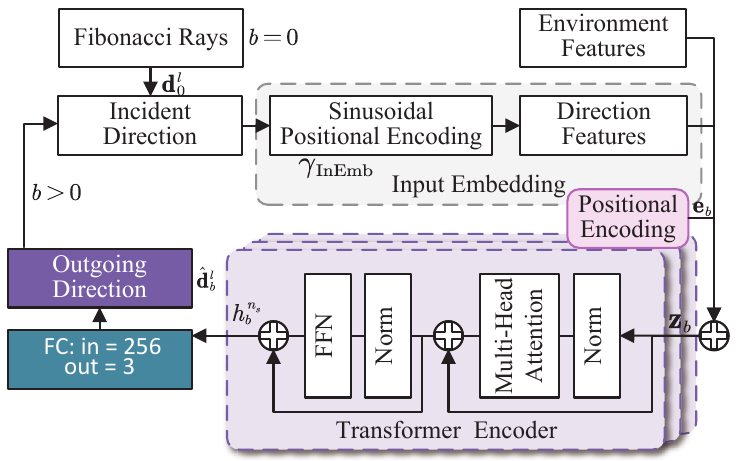}
	\caption{Structures of propagation path inference module. }
	\label{figure3}
\end{figure}
Structure of path inference module is illustrated in Fig.~\ref{figure3}.
This module is designed to sequentially infer directional evolution of rays.
At the initial propagation order $b=0$, incident direction $\mathbf{d}_0$ is initialized by uniformly sampling $L$ candidate directions on unit sphere using Fibonacci sequence.
Each sampled direction $\mathbf{d}_0^{l} \in \mathbb{R}^3$ is parameterized as
\begin{equation}
	\begin{aligned}
		\mathbf{d}_0^{l}
		=
		\Bigl(
		\sqrt{1 - z_l^2}\cos(2\pi l \Phi), 
		\sqrt{1 - z_l^2}\sin(2\pi l \Phi), 
		z_l
		\Bigr),\\
		z_l = 1 - \frac{2l}{L},
		\quad l = 1,\ldots,L .
	\end{aligned}
\end{equation}
where $\Phi = (\sqrt{5}-1)/2$ is inverse golden ratio.

At input embedding block, raw directional coordinates are not directly used as network inputs.
Instead, each predicted directional vector $\hat{\mathbf{d}}$ is mapped into a higher dimensional space using sinusoidal positional encoding,
\begin{equation}
	\begin{aligned}
		\gamma_{\mathrm{InEmb}}(\hat{\mathbf{d}}) =
		\Bigl[
		\sin(2\pi \hat{\mathbf{d}}),\;
		\cos(2\pi \hat{\mathbf{d}}),\;
		\ldots, \\
		\sin(2^{K}\pi \hat{\mathbf{d}}),\;
		\cos(2^{K}\pi \hat{\mathbf{d}})
		\Bigr]
	\end{aligned}
	\label{equ20}
\end{equation}
This encoding improves network’s sensitivity to grained directional changes.

The input to Transformer encoder is constructed by $\hat{\mathbf{d}}_b$ and $\gamma_{\mathrm{MLP}}\!\left(
\mathcal{F}^{(3)}
\right)$, which is given by
\begin{equation}
	\mathbf{z}_b =
	\left[
	\gamma_{\mathrm{InEmb}}(\mathbf{\hat{d}}_b),\;
	\gamma_{\mathrm{MLP}}\!\left(
	\mathcal{F}^{(3)}
	\right)
	\right]
	+ \mathbf{e}_b,
	\quad
	\mathbf{z}_b \in \mathbb{R}^{D},
\end{equation}
where $\mathbf{e}_b \in \mathbb{R}^{D}$ is learnable position embedding with $D$ dimensions.

Then, embedded features $\mathbf{z}_b$ are processed by a block composed of $N_s$ stacked Transformer encoder layers.
Each encoder layer consists of a multi-head self-attention module followed by a feed-forward network, with residual connections and layer normalization applied throughout.
The forward propagation of pre-norm Transformer encoder can be expressed as
\begin{equation}
	\mathbf{h}_b^{n_s'}
	=
	\mathbf{h}_b^{n_s-1}
	+
	\gamma_{\mathrm{MHA}}
	\!\left(
	\gamma_{\mathrm{Norm}}\!\left(\mathbf{h}_b^{n_s-1}\right)
	\right),
	n_s = 1,\ldots,N_s .
\end{equation}

\begin{equation}
	\mathbf{h}_b^{n_s}
	=
	\mathbf{h}_b^{n_s'}
	+
	\gamma_{\mathrm{MHA}}
	\!\left(
	\gamma_{\mathrm{FFN}}\!\left(\mathbf{h}_b^{n_s'}\right)
	\right),
	n_s = 1,\ldots,N_s .
\end{equation}
where $\mathbf{h}_b^{0} = \mathbf{z}_b$. After Transformer encoder, a fully connected layer mapping a 256-dim feature to a 3-dim outgoing ray direction $\mathbf{\hat{d}}_{b+1}^l$.

\subsubsection{Field Computation Stage}
This module takes environment features, current propagation state, and EM material parameters as joint inputs. Propagation features are constructed according to (\ref{equ20}), while EM material parameters are embedded into a high-dimensional latent space via a full connected layer to enhance their representational capacity. The resulting representations are concatenated along feature dimension and passed to MLP with a layer configuration of (128, 256, 68, 8). 
The output dimension is corresponding to the real and imaginary components of the polarimetric complex amplitude $\mathbf{I}$.
ReLU activation functions are employed throughout network.
Based on propagation directions predicted at path searching stage and $\mathbf{I}$, channel parameters of each multipath component can be subsequently computed.

In addition, a ray--scene interaction module determines whether a predicted ray reaches Rx. If a valid reception is detected, the propagation process is terminated; otherwise, network is iteratively invoked to generate next interaction until predefined maximum order is reached.

\subsection{Loss Functions}
\label{Loss function}
The loss function is composed of two terms: propagation direction accuracy and amplitude accuracy. These two parts correspond to path searching and field computation stage of RT simulation, respectively.

For path searching stage, propagation direction loss is defined based on cosine similarity between predicted and ground truth propagation directions, denoted by $\hat{\mathbf{d}} \in \mathbb{R}^3$ and $\mathbf{d} \in \mathbb{R}^3$, respectively. The direction loss for the $i$-th sample is formulated as
\begin{equation}
	\mathcal{L}_{\mathrm{dir}}
	=
	\frac{1}{N_{dir}}
	\sum_{i=1}^{N_{dir}}
	\left(
	1
	-
	\frac{
		\hat{\mathbf{d}}_i^{\mathsf{T}} \mathbf{d}_i
	}{
		\lVert \hat{\mathbf{d}}_i \rVert_2
		\lVert \mathbf{d}_i \rVert_2
	}
	\right),
\end{equation}
where $N_{dir}$ denotes total number of training samples.

For field computation stage, different loss formulations are adopted for deterministic and non-deterministic paths due to their power scales. Let $\hat{\mathbf{I}}_i$ and $\mathbf{I}_i$ denote predicted and ground truth polarimetric complex amplitude of the $i$-th path, respectively. For deterministic paths, which exhibit relatively large power values, attenuation loss is defined using mean squared error in linear domain as
\begin{equation}
	\mathcal{L}_{\mathrm{att}}^{\mathrm{det}}
	=
	\frac{1}{N_{\mathrm{det}}}
	\sum_{i=1}^{N_{\mathrm{det}}}
	\left(
	\hat{\mathbf{I}}_i - \mathbf{I}_i
	\right)^2,
\end{equation}
where $N_{\mathrm{det}}$ denotes number of deterministic paths.
By contrast, non-deterministic path power is typically on the order of $10^{-3}$.
To improve numerical stability and balance gradient magnitudes during training, attenuation loss for non-reflective paths is computed in logarithmic power domain as
\begin{equation}
	\mathcal{L}_{\mathrm{att}}^{\mathrm{non}}
	=
	\frac{1}{N_{\mathrm{non}}}
	\sum_{i=1}^{N_{\mathrm{non}}}
	\left(
	10 \log_{10} \hat{\mathbf{I}}_i^2
	-
	10 \log_{10} \mathbf{I}_i^2
	\right)^2,
\end{equation}
where $N_{\mathrm{non}}$ denotes number of non-deterministic paths.

The final training loss of the proposed network is obtained by combining the above components:
\begin{equation}
	\mathcal{L}
	=
	\lambda_1\mathcal{L}_{\mathrm{dir}}
	+
	\lambda_2\mathcal{L}_{\mathrm{att}},
	\label{equ27}
\end{equation}
where $\lambda_1$ and $\lambda_2$ are weighting coefficients that balance contributions of attenuation losses.

\section{Network Implementation}
\label{Network Implementation}
This chapter outlines experimental setup, including training algorithm, dataset composition, and measurement campaign.
\subsection{Training Algorithm}
A ray-level dataset is used for training and evaluation, where each sample corresponds to a single multipath component defined by Tx location $\mathbf{p}_{\mathrm{Tx}}$, Rx location $\mathbf{p}_{\mathrm{Rx}}$, the point cloud, $(\sigma, \varepsilon_r, S, K_x)$, and $(\mathbf{d}, \mathbf{I})$.

In each training iteration, a mini-batch of $T$ samples is randomly selected. For each sample, $\mathbf{d}_0$ is fed into network to determine the first interaction point $\mathbf{p}_0$. A local point cloud centered at $\mathbf{p}$ is encoded to extract environmental features. These features, together with incoming direction, are used to predict outgoing propagation direction $\hat{\mathbf{d}}$, while predicted direction, EM material parameters, and environment features are used to estimate  polarimetric complex amplitude $\hat{\mathbf{I}}$.
This procedure is iteratively repeated until the ray reaches $\mathbf{p}_{Rx}$ or the maximum propagation order $B$ is met. Overall procedure is summarized in Algorithm \ref{alg:pointnert}.
Model training is performed by minimizing loss function in (\ref{equ27}) using Adam optimizer with learning rate $\alpha$ and momentum parameters $\beta_1$ and $\beta_2$.

\begin{algorithm}[t]
	\caption{Network Training and Testing}
	\label{alg:pointnert}
	
	\textbf{Data:} $\mathrm{PointCloud}$, $(\sigma, \varepsilon_r, S, K_x)$,
	$\mathbf{p}_{\mathrm{Tx}}, \mathbf{p}_{\mathrm{Rx}}$,
	$(\mathbf{p}, \mathbf{d}, \mathbf{I})$ \;
	\textbf{Hyperparameters:} $(\beta_1, \beta_2)$, $(\lambda_1, \lambda_2)$, $B$\;
	\textbf{Initialize:} network parameters $\Omega$ \;
	
	\vspace{0.3em}
	\textbf{Training Stage (Single-Bounce)}
	
	\While{not converged}{
		Sample a mini-batch of $T$ rays $\{\mathbf{d}_0^t\}_{t=1}^{T}$ \;
		\For{$t = 1$ \KwTo $T$}{
			$\mathbf{p}_0^t = \mathrm{Intersect}(\mathrm{PointCloud}, \mathbf{p}_{\mathrm{Tx}}, \mathbf{d}_0^t)$ \;
			$\mathcal{S}_0^t = \mathrm{Crop}(\mathrm{PointCloud}, \mathbf{p}_0^t)$ \;
			$\mathbf{f}_0^t = \mathrm{SceneEnc}(\mathcal{S}_0^t)$ \;
			$\hat{\mathbf{d}}_1^t = \mathrm{DirPred}(\mathbf{f}_0^t, \mathbf{d}_0^t)$ \;
			$\hat{\mathbf{I}}_1^t = \mathrm{AmpPred}(\mathbf{f}_0^t, \hat{\mathbf{d}}_1^t, \mathcal{M})$ \;
		}
		$\mathcal{L} = \mathrm{Loss}(\hat{\mathbf{d}}_1, \hat{\mathbf{I}}_1; \mathbf{d_1}, \mathbf{I_1})$ \;
		$\Omega \leftarrow \mathrm{Adam}(\Omega, \nabla_{\Omega}\mathcal{L})$ \;
	}
	
	\vspace{0.3em}
	\textbf{Testing Stage (Multi-Bounce)}
	
	\For{each launch direction $\mathbf{d}_0$}{
		$\mathbf{p}_0 = \mathrm{Intersect}(\mathrm{PointCloud}, \mathbf{p}_{\mathrm{Tx}}, \mathbf{d}_0)$ \;
		$b \leftarrow 0$ \;
		\While{$b < B$}{
			$\mathcal{S}_b = \mathrm{Crop}(\mathrm{PointCloud}, \mathbf{p}_b)$ \;
			$\mathbf{f}_b = \mathrm{SceneEnc}(\mathcal{S}_b)$ \;
			$\hat{\mathbf{d}}_{b+1} = \mathrm{DirPred}(\mathbf{f}_b, \mathbf{d}_b)$ \;
			$\hat{\mathbf{p}}_{b+1} = \mathrm{Intersect}(\mathcal{S}, \mathbf{p}_b, \hat{\mathbf{d}}_{b+1})$ \;
			$\hat{\mathbf{I}}_{b+1} = \mathrm{AmpPred}(\mathbf{f}_b, \mathbf{d}_b, \hat{\mathbf{d}}_{b+1}, \mathcal{M})$ \;
			\If{$\mathrm{HitRx}(\hat{\mathbf{p}}_{b+1})$}{
				Store this path and \textbf{break} \;
			}
			$\mathbf{p}_b \leftarrow \hat{\mathbf{p}}_{b+1}$,
			$\mathbf{d}_b \leftarrow \hat{\mathbf{d}}_{b+1}$,
			$b \leftarrow b+1$ \;
		}
	}
\end{algorithm}
Since different orders within a multipath are mutually independent, only single-bounce rays are used during training stage to ensure stable convergence of network. At test stage, trained model is applied in an iterative manner to generate multi-bounce propagation paths, enabling efficient reconstruction of complex multipath channels. The same propagation configuration in Section~\ref{Path tracing} is adopted to ensure consistency between dataset and learning. The detailed training parameter configuration is shown in Table \ref{tab:train_hparams}.
\begin{table}[h]
	\centering
	\caption{Training Hyperparameters of PointNeRT}
	\label{tab:train_hparams}
	\begin{tabular}{ll}
		\toprule
		\textbf{Hyperparameter} & \textbf{Value} \\
		\midrule
		Optimizer & Adam $(0.9, 0.999)$ \\
		Learning rate & $0.0001$ \\
		Batch size & 64 \\
		Training epochs & 800 \\
		Weight decay & 0.8 \\
		\midrule
		Loss weights (deterministic paths) & $(\lambda_1,\lambda_2) = (1,\;5)$ \\
		Loss weights (non-deterministic paths) & $(\lambda_1,\lambda_2) = (1,\;0.001)$ \\
		\bottomrule
	\end{tabular}
\end{table}

\subsection{Data Preparation}
\label{Data Preparation}
Multipath dataset is generated through calibrated RT simulator based on measured point clouds, with detailed RT configuration described in Section~\ref{Path tracing}. All simulations are carried out at 28~GHz.
To emphasize propagation components with significant contributions to signal, a power-based filtering is applied.
Specifically, only multipath components within 40~dB below maximum received power of each snapshot are retained.

The base dataset is constructed from scenario Room1. In this environment, 420 Tx–Rx pairs are deployed, and a total of 9,343 valid multipath components are obtained. The resulting dataset is randomly partitioned into training and testing sets with a ratio of 5:1. To focus on NLOS path modeling, LOS components are excluded from both sets.

In addition to dataset constructed from Room1, two additional scenarios with distinct structural characteristics, namely Room2 and Room3, are introduced to evaluate generalization capability of network. In these scenarios, 40 and 30 Tx--Rx pairs are deployed, respectively.
In Room2, overall room geometry remains unchanged. Locations of key scattering objects (e.g., columns) are varied to emulate dynamic environments. This configuration contains 918 valid multipath components.
Room3 represents a structurally different room layout with a total of 1,038 multipath components, and is used to evaluate network’s robustness and predictive performance under significant geometric variations. The structural layouts of three rooms are illustrated in Fig.~\ref{figure4}.

\subsection{Scenario and Measurements}
\label{Scenario and Measurements}
\begin{figure}[!tb]
	\centering
	\subfloat[]{\label{fig:fig4-b}
		\includegraphics[width=3.2in]{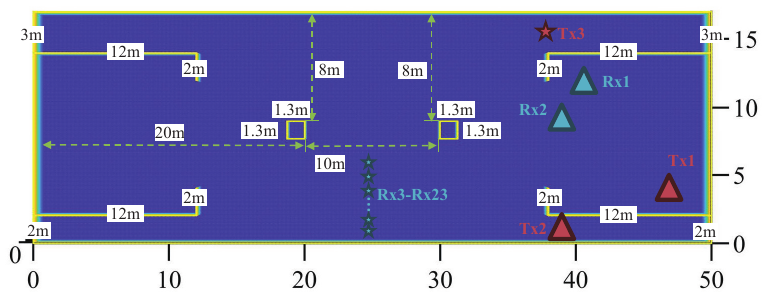}}\\[-1pt]
	\subfloat[]{\label{fig:fig4-c}
		\includegraphics[width=1.75in]{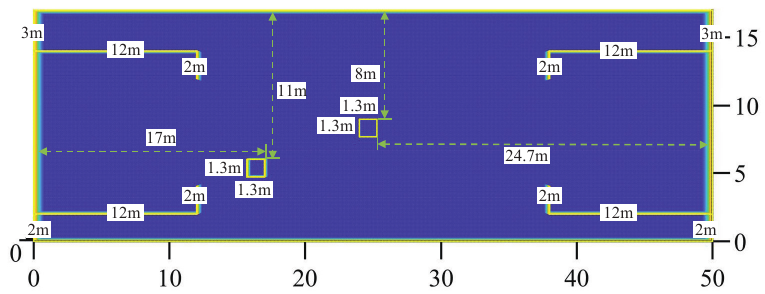}}
	\subfloat[]{\label{fig:fig4-d}
		\includegraphics[width=1.75in]{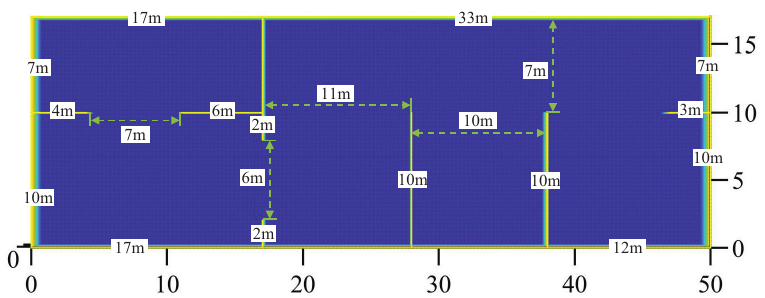}}
	\caption{Floor plan overlaid with point clouds. Ceiling of point clouds have been removed for a better visualization. Pentagrams mark measurement links used for RT calibration, while triangles denote evaluation links for assessing performance of proposed network. (a) Measurement environment layout (Room1). (b) Room with relocated key scatterers (Room2). (c) Room with a modified layout (Room3).}
	\label{figure4}
\end{figure}
\begin{figure}[!tb]
	\centering
	\includegraphics[width=2.5in]{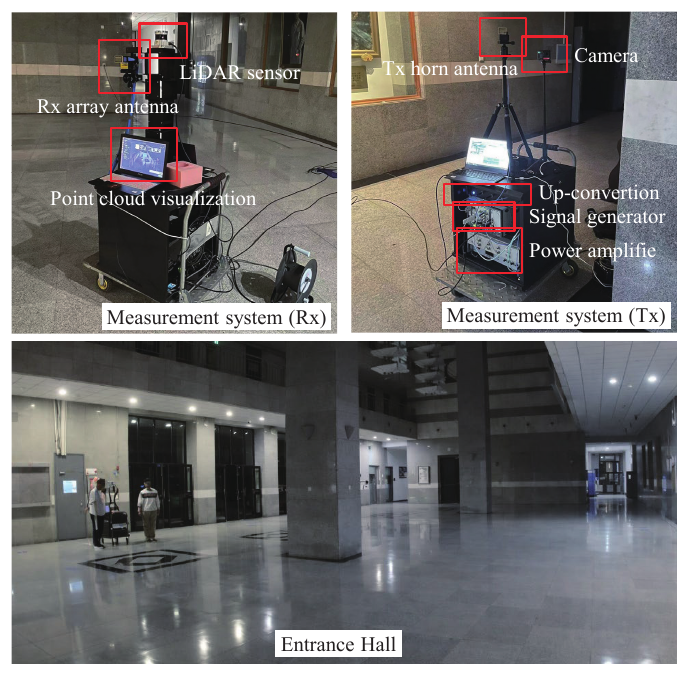}
	\caption{Measurement environment and on-site channel measurement system.}
	\label{figure9}
\end{figure}

A real-world 3D point cloud and channel measurement data were acquired through an on-site measurement campaign carried out at an entrance hall in Beijing Jiaotong University, as shown in Fig.~\ref{figure9}.
The entrance hall has approximate dimensions of $50 \times 17 \times 4.5~\mathrm{m}^3$ and contains large smooth wall and window surfaces, along with typical scattering objects such as structural columns.
In addition, solid walls separate the main hall from an adjacent entrance area, thereby supporting both LOS and NLOS propagation conditions.
To accurately capture structure model, measurement area was scanned with OS1 laser scanner equipped with 1024 laser beams. It densely samples surrounding space by measuring distances to discrete points, which are saved as points in $x$-, $y$-, and $z$-coordinates. The resulting point cloud consists of approximately $1.96$ million points, with an average separation distance of about $8~\mathrm{cm}$. The reconstructed point cloud is illustrated in Fig. \ref{figure4}\subref{fig:fig4-b}.

To enable accurate calibration of RT simulator, an on-site channel measurement campaign were conducted at a center frequency of $28~\mathrm{GHz}$ with a signal bandwidth of $1~\mathrm{GHz}$. Channel frequency responses over 1024 OFDM subcarriers were captured at fixed measurement positions from a single-antenna horn transmitter to a planar 4$\times$8 antenna array with $\lambda/2$ spacing and vertically polarized elements \cite{ref35}. The measured channel data collected at Tx3 and Rx3–Rx23 in Fig. \ref{figure4}\subref{fig:fig4-b} were used to calibrate EM material parameters in RT simulator before data preparation.

\section{Experiments and Performance Evaluation}
\label{Experiments and Performance Evaluation}
This section presents experimental analysis of PointNeRT.
We first evaluate overall performance of the proposed approach. 
Next, we assess model's generalization ability.
Finally, the impact of different input feature combinations on prediction accuracy is investigated.
Throughout this section, ``ground truth" refers to RT simulated results generated by calibrated RT simulator.

\subsection{Performance Validation in Room1}
\label{Accuracy Validation}

\begin{figure*}[!tb]
	\centering
	\makebox[\textwidth][c]{%
		\subfloat[]{\label{fig:fig5-a}
			\includegraphics[width=1.7in]{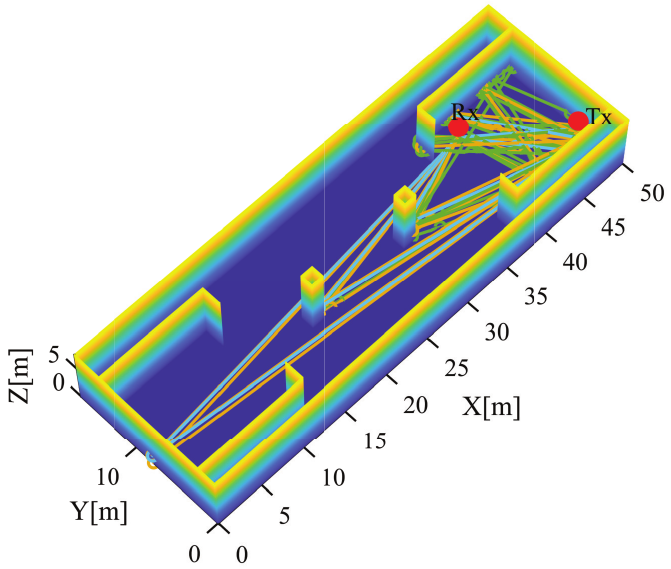}}
		\subfloat[]{\label{fig:fig5-b}
			\includegraphics[width=1.7in]{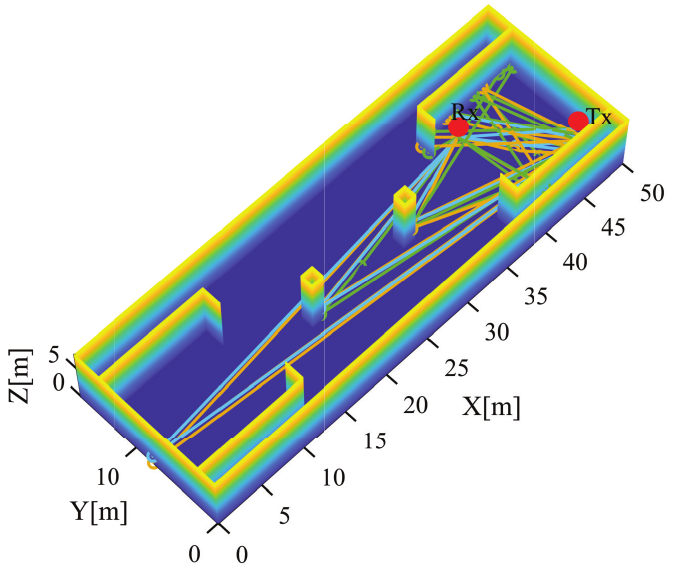}}
		\subfloat[]{\label{fig:fig5-c}
			\includegraphics[width=1.7in]{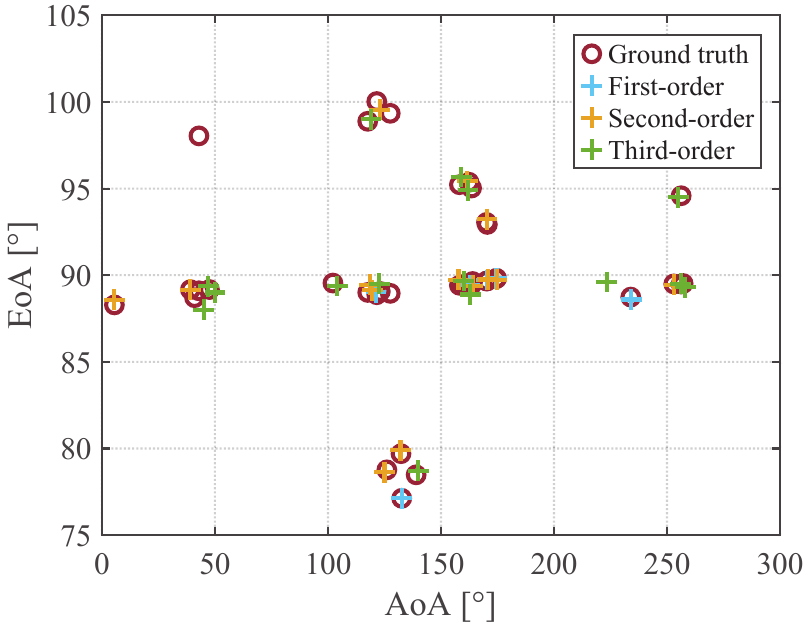}}
		\subfloat[]{\label{fig:fig5-d}
			\includegraphics[width=1.7in]{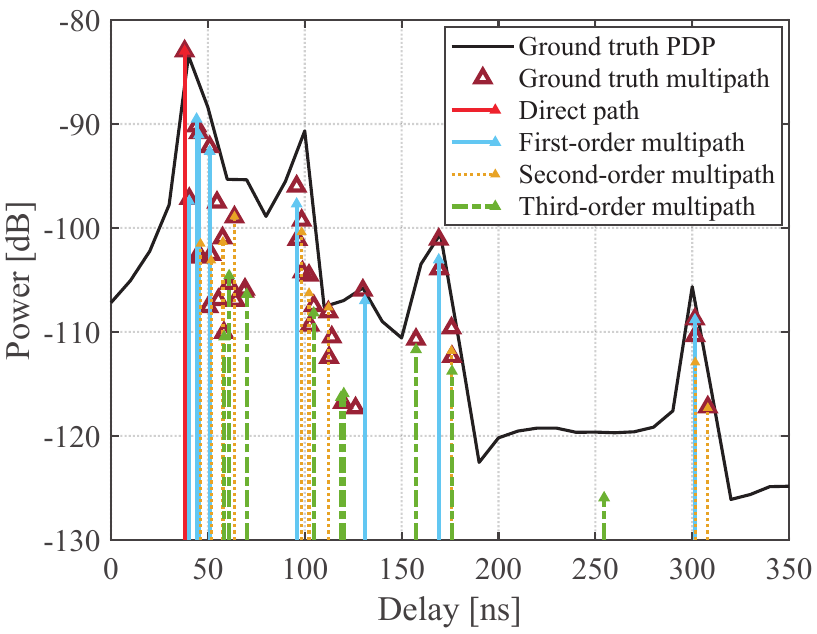}}
	}
	\vspace{-0.05in}
	\makebox[\textwidth][c]{%
		\subfloat[]{\label{fig:fig5-e}
			\includegraphics[width=1.7in]{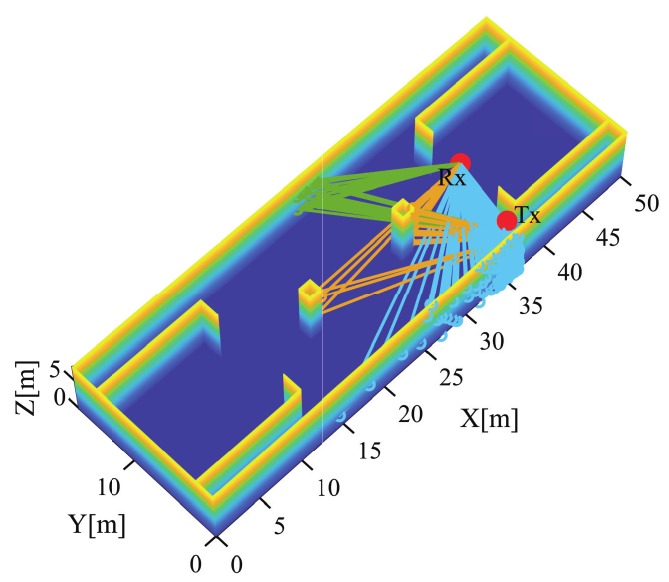}}
		\subfloat[]{\label{fig:fig5-f}
			\includegraphics[width=1.7in]{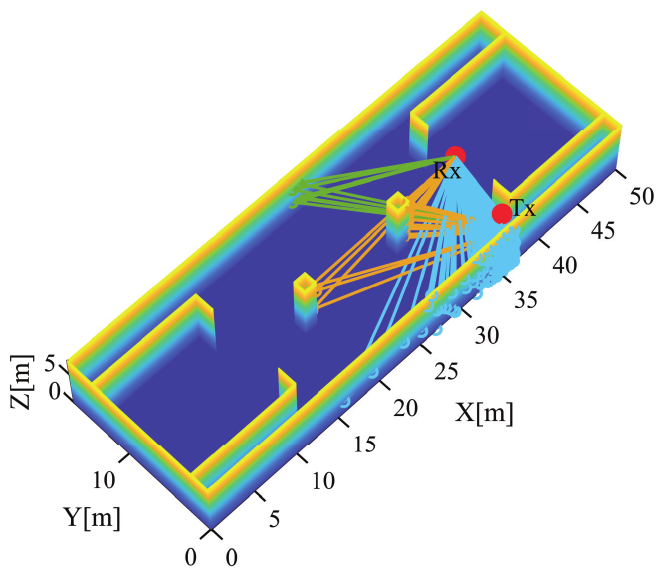}}
		\subfloat[]{\label{fig:fig5-g}
			\includegraphics[width=1.7in]{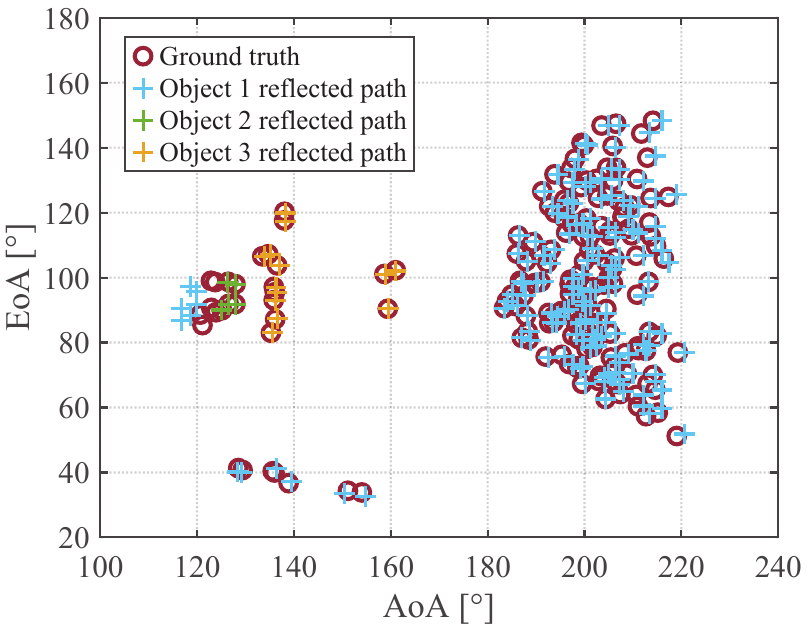}}
		\subfloat[]{\label{fig:fig5-h}
			\includegraphics[width=1.7in]{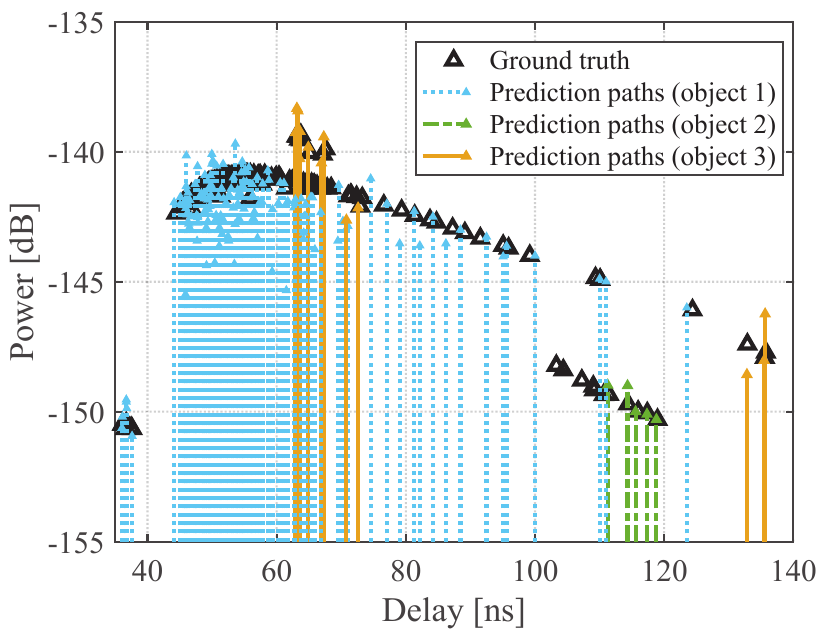}}			
	}
	\caption{Multipath comparison between ground truth and PointNeRT: (a)–(d) LOS case for Tx1–Rx1 (colors denote bounce order) and (e)–(h) NLOS case for Tx2–Rx2 (colors denote scattering sources). Columns show ground truth, prediction, azimuth comparison, PDP comparison from left to right.}
	\label{figure5}
\end{figure*}
Accuracy validation of PointNeRT consists of two components: path prediction and field computation. 
Fig. \ref{figure5}\subref{fig:fig5-a}-\subref{fig:fig5-c} present multipath prediction results for Tx1--Rx1 link under a LOS scenario. In this environment, propagation is dominated by deterministic paths. Comparison of azimuth of arrival (AOA) and elevation of arrival (EOA) is illustrated in Fig.~\ref{figure5}\subref{fig:fig5-c}. It shows that PointNeRT achieves good agreement, especially for first- and second-order paths.
As bounce order increases, errors accumulate during hop-by-hop prediction process, resulting in deviations in both angle and spatial position for higher-order multipath components.
In NLOS conditions, propagation is mainly characterized by non-deterministic paths. The results in Fig.~\ref{figure5}\subref{fig:fig5-e}--\subref{fig:fig5-g} indicate that multipath components scattered by different objects are accurately predicted by PointNeRT.

Field computation validation is shown in Fig. \ref{figure5}\subref{fig:fig5-d}\subref{fig:fig5-h}. 
Fig. \ref{figure5}\subref{fig:fig5-d} compares power delay profile (PDP) for Tx1.
It can be observed that multipath components predicted by PointNeRT align well with dominant energy peaks of ground truth.
Both path delays and power distributions show good agreement, allowing overall PDP shape and propagation characteristics to be accurately reconstructed. While a small number of third-order reflections are not fully recovered, their impact on dominant energy distribution is limited.
Similarly, PDP comparison for Tx2 in Fig.~\ref{figure5}\subref{fig:fig5-h} shows that PointNeRT successfully captures power distribution trends of multipath components scattered by different objects, reflecting its ability to implicitly learn EM propagation characteristics. 
\begin{figure}[!tb]
	\centering
	\includegraphics[width=3.2in]{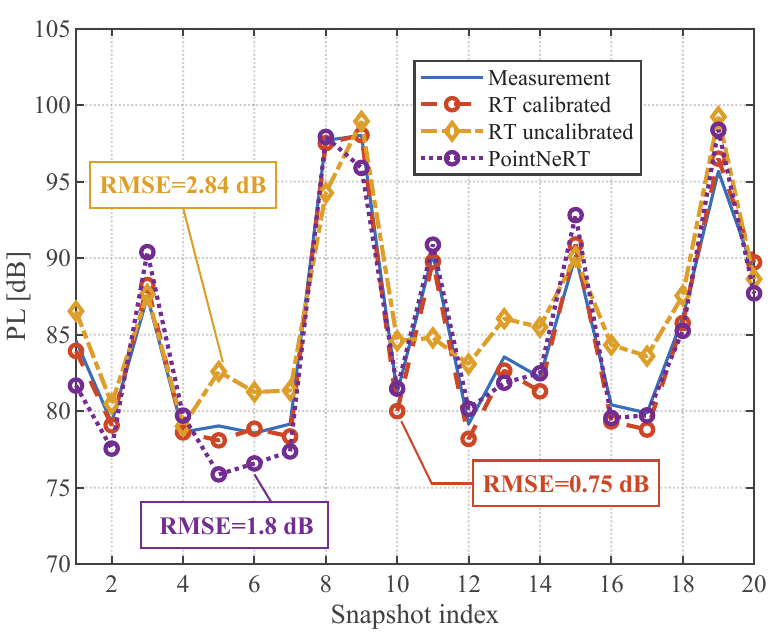}
	\caption{Predicted PL by various methods compared to measurements. RMSE is calculated relative to measured data.}
	\label{figure8}
\end{figure}
We next evaluate field computation of PointNeRT by comparing predictions with measured data in Fig. \ref{figure8}. RT simulator used for dataset generation shows a 2.09 dB improvement in accuracy after calibration. Additionally, error between predicted and calibrated RT simulator is 1.60 dB.
The validation result implies predict model presented herein is well suited for simulating channel data for stochastic channel modeling purposes.

In terms of computational efficiency, the proposed network avoids scene reconstruction step required by mesh-based RT, which involves generating a mesh from point cloud for mapping. This reconstruction process takes approximately 20.05 hours for Room 1.
In addition, completing simulations of 100 rays require 1.65 s using the proposed approach, whereas point-based RT method in \cite{ref24} takes 2.81 s per path. The computational efficiency can be further improved through multi-threaded acceleration.

\subsection{Generalization Evaluation in Room2 and Room3}
\label{Generalization Performance Evaluation}

In this subsection, PL and root-mean-square delay spread (DS) are used as condensed parameters to evaluate generalization performance of PointNeRT in Room 2 and Room 3.

\begin{figure}[!tb]
	\centering
	\subfloat[]{\label{fig:fig6-a}
		\includegraphics[width=1.7in]{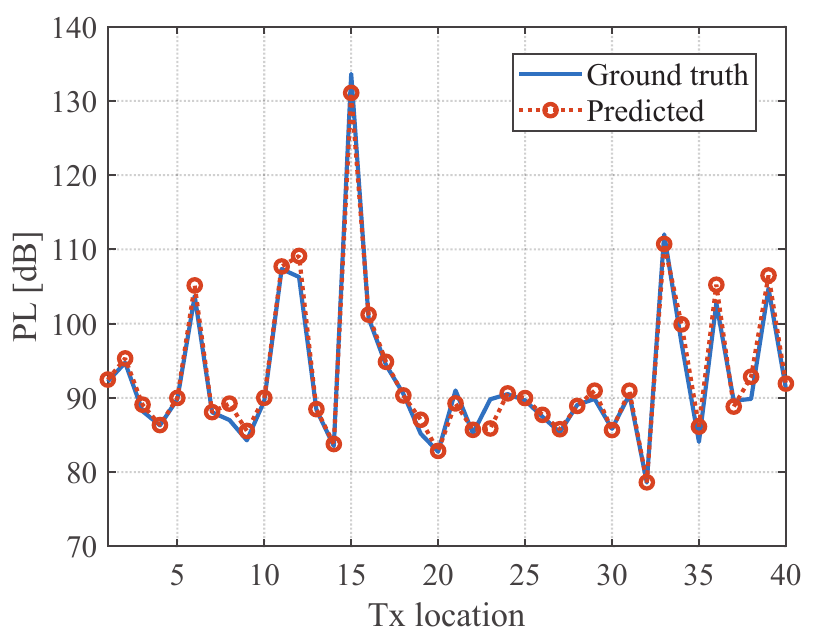}}
	\subfloat[]{\label{fig:fig6-b}
		\includegraphics[width=1.66in]{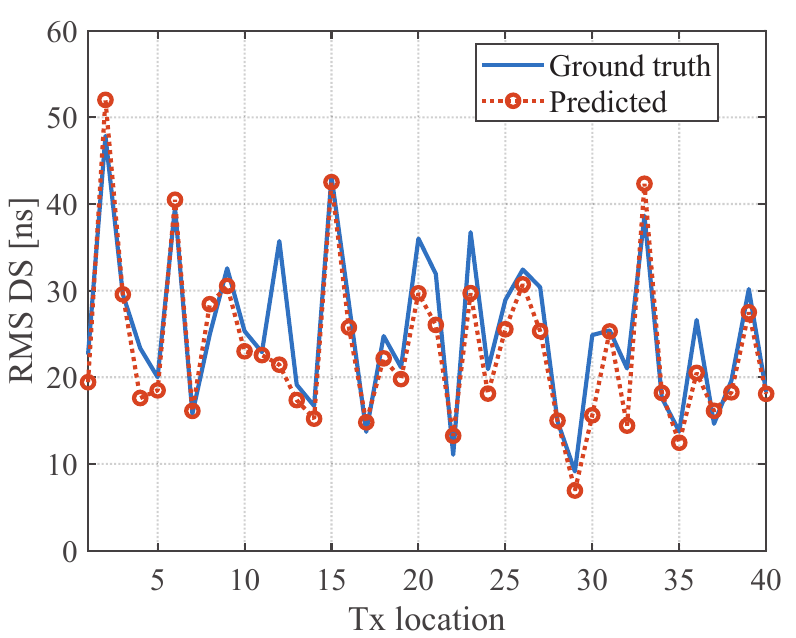}}
	\caption{Comparison of condensed parameters of multipath radio channels obtained from ground truth and predicted in Room2.(a) PL. RMSE = 1.71 dB. (b) DS. RMSE = 4.15 ns.}
	\label{figure6}
\end{figure}

\begin{figure}[!t]
	\centering
	\subfloat[]{\label{fig:fig7-a}
		\includegraphics[width=1.7in]{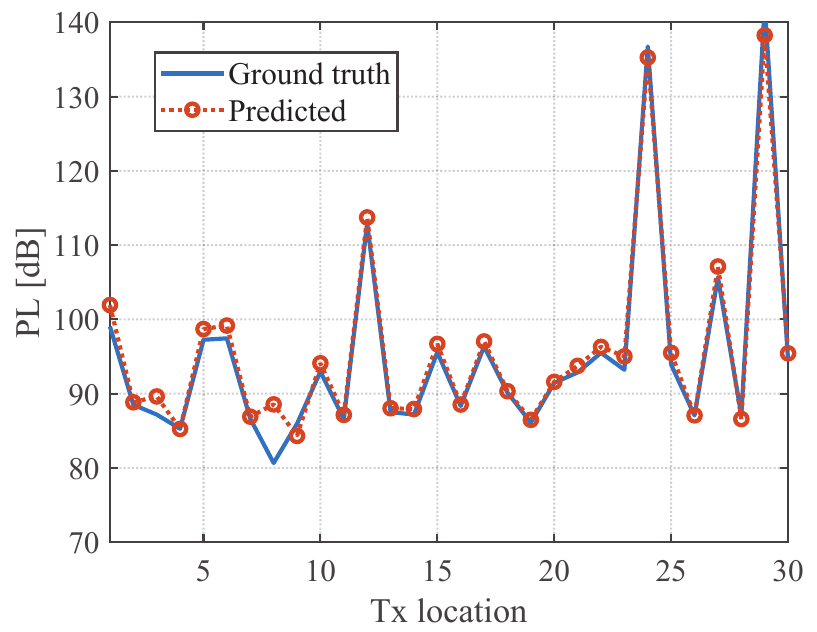}}
	\subfloat[]{\label{fig:fig7-b}
		\includegraphics[width=1.66in]{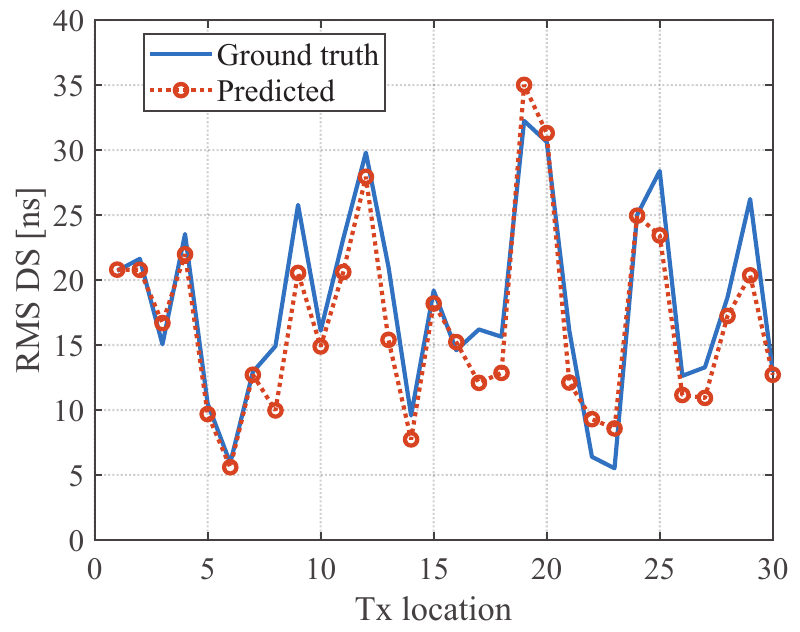}}
	\caption{Comparison of condensed parameters of multipath radio channels obtained from ground truth and predicted in Room3.(a) PL. RMSE = 1.98 dB. (b) DS. RMSE = 2.84 ns.}
	\label{figure7}
\end{figure}

Comparisons of channel parameters in different scene are presented in Fig.~\ref{figure6} and Fig.~\ref{figure7}, respectively. 
Predicted PL and DS generally follow ground truth trends with reasonable accuracy. The results confirm robust generalization of proposed model in unseen environments. 
However, several local peaks in measured curves are not fully reproduced.
A representative discrepancy can be observed at Tx12 in Room2, where predicted DS is approximately 14~ns lower than ground truth value. 
Further analysis shows that this error originates from a third-order reflection path reflected by a column. 
Specifically, the third interaction occurs on column surface, and due to accumulated angular deviations in hop-by-hop prediction process, predicted reflection direction at the third bounce deviates by approximately $5^\circ$. 
Given limited reflecting area of column, this deviation prevents ray from intersecting scatterer, leading to the absence of corresponding reflected path at Rx.
In addition, prediction accuracy of condensed channel parameters is slightly lower in NLOS than in LOS cases. Relative contribution of second- and third-order reflections becomes significant in NLOS. This is also clearly given in Table \ref{tab2}. Although third-order multipath exhibit comparable power levels in both LOS and NLOS links, their contribution to power and delay metrics is different when a LOS component exists, whereas they can account for more than 10\% of received power once direct path is blocked.
\begin{table*}[!t]
	\centering
	\caption{Average power contributions of different order propagation paths and RMSE of predicted power at each Rx \\in Room2 and Room3 (Average  received Power Levels Shown in Brackets)}
	\label{tab2}
	\setlength{\tabcolsep}{6pt}
	\renewcommand{\arraystretch}{1.2}
	
	\begin{tabular}{ccccccc}
		\toprule
		\textbf{} 
		& \textbf{} 
		& \textbf{\shortstack{Link\\type}}
		& \raisebox{1.1ex}{\textbf{Direct path} }
		& \textbf{\shortstack{First-order\\multipath}} 
		& \textbf{\shortstack{Second-order\\multipath}} 
		& \textbf{\shortstack{Third-order\\multipath}} \\
		\midrule
		
		\multirow{4}{*}{%
			\shortstack{Average power\\contributions}%
			\hspace{0.5em}%
			\raisebox{-4ex}{\vrule width 0.5pt height 13ex}%
		}
		& \multicolumn{1}{c|}{\multirow{2}{*}{\shortstack{Ground\\truth}}}
		& LOS  & 80.18\% (-79.97 dB) & 16.08\% (-92.69 dB) & 3.14\% (-102.45 dB) & 0.60\% (-109.56 dB) \\
		& \multicolumn{1}{c|}{}
		& NLOS & 0\% (N/A)        & 60.95\% (-104.76 dB) & 28.81\% (-104.62 dB) & 10.24\% (-111.84 dB) \\
		
		\cmidrule(l{7.5em}){1-7}
		
		& \multicolumn{1}{c|}{\multirow{2}{*}{Predicted}}
		& LOS  & 82.07\% (-79.97 dB) & 15.18\% (-92.91 dB) & 2.51\% (-101.95 dB) & 0.24\% (-109.32 dB) \\
		& \multicolumn{1}{c|}{}
		& NLOS & 0\% (N/A)        & 65.96\% (-104.83 dB) & 31.42\% (-103.03 dB) & 2.62\% (-113.27 dB) \\
		
		\midrule
		
		\multicolumn{2}{c|}{RMSE of predicted power [dB]}
		& -- & 0 & 0.22 & 1.32 & 4.03 \\
		
		\multicolumn{2}{c|}{Multipath angular error [$^\circ$]}
		& -- & 0 & 1.74 & 3.28 & 6.56 \\
		
		\bottomrule
	\end{tabular}
\end{table*}

Overall, the results demonstrate that PointNeRT exhibits strong generalization capability while maintaining reliable accuracy in different environments. 
This property is particularly appealing for simulation scenarios involving dynamic environments (e.g., moving vehicles or reconfigurable objects). 

\subsection{Impact of Environmental Feature Fusion on Prediction Performance}
\label{Impact of Environmental Feature Fusion on Prediction Performance}
According to EM waves propagation mechanisms, surface normal vectors provide physical constraints on propagation directions, controlling spatial evolution of rays, while material information characterizes intrinsic EM properties of interaction surfaces and affects multipath attenuation. Motivated by these physical insights, different combinations of environmental features are designed as network inputs to evaluate their contributions to multipath prediction performance.
Four input configurations are therefore considered to analyze influence of different environmental features: point cloud geometry only (Env.), point cloud with interaction-point normal vectors (Env. + Norm.), point cloud with EM material parameters (Env. + Mat.), and the joint use of point cloud, normals vectors, and EM material parameters (Env. + Norm. + Mat.).
Specifically, surface normals are provided only to path searching stage, while EM material parameters are introduced in field computation stage, since their distinct roles in geometric path determination and EM attenuation predicting.

\begin{table}[!t]
	\centering
	\caption{RMSE Between Ground truth and Predicted Large-Scale Channel Parameters with Different Input Features}
	\label{tab:scattering_params}
	\setlength{\tabcolsep}{3pt} 
	\begin{tabular}{lcccccc}
		\toprule
		\multirow{2}{*}{\textbf{Input Features}} 
		& \multicolumn{3}{c}{\textbf{PL [dB]}} 
		& \multicolumn{3}{c}{\textbf{DS [ns]}} \\
		\cmidrule(lr){2-4} \cmidrule(lr){5-7}
		& Room1 & Room2 & Room3 
		& Room1 & Room2 & Room3 \\
		\midrule
		Env.
		& 2.56 & 5.79 & 4.16 
		& 4.59  & 8.17  & 3.56\\
		
		Env. + Norm.
		& 2.31 & 5.49 & 3.94
		& 4.41  & 7.93  & 3.53\\
		
		Env. + Mat.
		& \textbf{1.59} & \textbf{1.71} & \textbf{1.98}
		& \textbf{3.19}  & \textbf{4.15}  & 2.84 \\
		
		Env. + Nor. + Mat.
		& 1.95 & 2.20 & 2.28
		& 3.83  & 4.59  & \textbf{2.48}\\
		\bottomrule
	\end{tabular}
\end{table}

The root mean square error (RMSE) between ground truth and predicted large-scale channel parameters under different input features is summarized in Table~\ref{tab:scattering_params}.
Using Env. as baseline, inclusion of surface normals leads to limited improvements. Specifically, PL error decreases by 0.25~dB, 0.30~dB, and 0.22~dB in Room1, Room2, and Room3, respectively, while corresponding DS error reductions are 0.18~ns, 0.24~ns, and 0.03~ns. 
This observation suggests that network can implicitly learn surface structural characteristics directly from point cloud.
In contrast, incorporating EM material parameters consistently improves prediction accuracy, with PL error reductions of 0.97~dB, 4.08~dB, and 2.18~dB in three rooms, corresponding to relative improvements exceeding 35\%. 
Notably, larger gains observed in generalization scenarios compared to Room 1 suggest that network is able to implicitly learn EM material  properties from geometric cues, while explicit material information further enhances robustness and accuracy when extrapolating to heterogeneous environments.
Similar performance gains are observed for DS.
This improvement can be attributed to the fact that EM material parameters provide additional physical constraints beyond geometry, yielding improved performance when room structures are similar. 
In contrast, jointly incorporating normals offers limited benefit and may introduce redundancy once geometric and EM material information is available.
In addition, comparison shows that prediction errors in Room2 are generally higher than in Room3. 
Room3 is dominated by planar structures and deterministic first-order reflections, which are easier to predict. In contrast, Room2 has a more open layout with varying column positions, resulting in a higher proportion of non-deterministic paths and increased multipath complexity.

\section{Conclusion}
\label{Conclusion}
In this paper, we propose a physics aware neural RT surrogate named PointNeRT for point-based propagation channel simulation. The proposed framework integrates physics aware interaction modeling with a carefully designed loss formulation, enabling network to learn propagation behaviors that are consistent with EM principles. By dividing propagation mechanisms into deterministic and non-deterministic interactions and modeling them using different neural networks, physical constraints are explicitly embedded into learning process. We implement our prediction method in a real entrance hall and validate in three rooms with distinct geometric layouts. 
Results show that the PointNeRT achieves a favorable balance between prediction accuracy and computational efficiency, while exhibiting robust generalization to significant structural variations.
Furthermore, analysis of different input feature combinations shows that, although network can implicitly capture certain environmental surface characteristics and material properties, explicit material properties play a crucial role in enhancing generalization across heterogeneous scenarios.


\bibliographystyle{IEEEtran}  
\bibliography{reference}

@ARTICLE{ref1,
  author={Hoydis, Jakob and Aoudia, Fayçal Aït and Cammerer, Sebastian and Euchner, Florian and Nimier-David, Merlin and Brink, Stephan Ten and Keller, Alexander},
  journal={IEEE Trans. Mach. Learn. Commun. Networking},
  title={Learning Radio Environments by Differentiable Ray Tracing},
  year={2024},
  volume={2},
  number={},
  pages={1527-1539},
  doi={10.1109/TMLCN.2024.3474639}}

@ARTICLE{ref2,
      title={{6G White Paper on Localization and Sensing}},
      journal={University of Oulu: 6G Research Visions},
      author={Andre Bourdoux and Andre Noll Barreto and Barend van Liempd and Carlos de Lima and Davide Dardari and Didier Belot and Elana-Simona Lohan and Gonzalo Seco-Granados and Hadi Sarieddeen and Henk Wymeersch and Jaakko Suutala and Jani Saloranta and Maxime Guillaud and Minna Isomursu and Mikko Valkama and Muhammad Reza Kahar Aziz and Rafael Berkvens and Tachporn Sanguanpuak and Tommy Svensson and Yang Miao},
      year={Jun. 2020}
}

@ARTICLE{ref3,
  author={Studer, Christoph and Medjkouh, SaïD and Gonultaş, Emre and Goldstein, Tom and Tirkkonen, Olav},
  journal={IEEE Access},
  title={Channel Charting: Locating Users Within the Radio Environment Using Channel State Information},
  year={2018},
  volume={6},
  number={},
  pages={47682-47698},
  doi={10.1109/ACCESS.2018.2866979}}

@INPROCEEDINGS{ref4,
  author={Lodato, Francesca and Garzia, Andrea and Valbonesi, Simona and Ruello, Giuseppe and Iodice, Antonio and Matera, Francesco and Salvo, Pierpaolo and Massa, Rita},
  booktitle={Proc. IEEE Int. Symp. Meas. Netw. (M\&N)},
  title={Ray Tracing Tools Assessment for the Evaluation of {EMF} Levels Generated by {5G NR} Systems: An Overview},
  year={2024},
  volume={},
  number={},
  pages={1-6},
  doi={10.1109/MN60932.2024.10615802}}

@ARTICLE{ref5,
  author={He, Danping and Guan, Ke and Yan, Dong and Yi, Haofan and Zhang, Zhao and Wang, Xiping and Zhong, Zhangdui and Zorba, Nizar},
  journal={IEEE J. Sel. Areas Commun.},
  title={Physics and {AI}-Based Digital Twin of Multi-Spectrum Propagation Characteristics for Communication and Sensing in {6G} and Beyond},
  year={2023},
  volume={41},
  number={11},
  pages={3461-3473},
  doi={10.1109/JSAC.2023.3310108}}

@INPROCEEDINGS{ref6,
  author={Järveläinen, Jan and Kurkela, Matti and Karttunen, Aki and Haneda, Katsuyuki and Putkonen, Jyri},
  booktitle={Proc. Loughborough Antennas Propag. Conf. (LAPC)},
  title={70 {GHz} radio wave propagation prediction in a large office},
  year={2014},
  volume={},
  number={},
  pages={420-424},
  doi={10.1109/LAPC.2014.6996414}}

@article{ref7,
author = {Le Yu and Peng Gong},
title = {Google Earth as a virtual globe tool for Earth science applications at the global scale: progress and perspectives},
journal = {Int. J. Remote Sens.},
volume = {33},
number = {12},
pages = {3966--3986},
year = {2012},
publisher = {Taylor \& Francis},
doi = {10.1080/01431161.2011.636081}}

@ARTICLE{ref8,
  author={Koivumäki, Pasi and Steinböck, Gerhard and Haneda, Katsuyuki},
  journal={IEEE Trans. Antennas Propag.},
  title={Impacts of Point Cloud Modeling on the Accuracy of Ray-Based Multipath Propagation Simulations},
  year={2021},
  volume={69},
  number={8},
  pages={4737-4747},
  doi={10.1109/TAP.2021.3050482}}

@inproceedings{ref15,
  title={Gsrf: Complex-valued 3d gaussian splatting for efficient radio-frequency data synthesis},
  author={Yang, Kang and Dong, Gaofeng and Du, Wan and Srivastava, Mani and others},
  booktitle={Proc. The Thirty-ninth Annual Conference on Neural Information Processing Systems},
  year={2025}
}

@ARTICLE{ref16,
  author={Yang, Kaiqiao and Liu, Che and Yu, Wenming and Jun Cui, Tie},
  journal={IEEE Trans. Antennas Propag.},
  title={{PointEMRay}: A Novel Efficient {SBR} Framework on Point-Based Geometry},
  year={2025},
  volume={73},
  number={1},
  pages={375-390},
  doi={10.1109/TAP.2024.3502912}}

@article{ref17,
  title={Scene representation networks: Continuous 3d-structure-aware neural scene representations},
  author={Sitzmann, Vincent and Zollh{\"o}fer, Michael and Wetzstein, Gordon},
  journal={Advances in Proc. Neural Inf. Process. Syst.},
  volume={32},
  year={2019}
}

@article{ref18,
  title={Implicit neural representations with periodic activation functions},
  author={Sitzmann, Vincent and Martel, Julien and Bergman, Alexander and Lindell, David and Wetzstein, Gordon},
  journal={Advances in Proc. Neural Inf. Process. Syst.},
  volume={33},
  pages={7462--7473},
  year={2020}
}

@article{ref19,
title={Sionna {RT}: Technical Report},
author={{Aït Aoudia}, Fayçal and Hoydis, Jakob and Nimier-David, Merlin and Nicolet, Baptiste and Cammerer, Sebastian and Keller, Alexander},
year = {2025},
month = {Nov},
journal = {arXiv:2504.21719}
}

@article{ref20,
  title={Introduction to the Uniform Geometrical Theory of Diffraction},
  author={ Mcnamara, D.  and  Pistorius, C.  and  Malherbe, J. },
  year={1990}
}

@ARTICLE{ref21,
  author={Degli-Esposti, Vittorio and Kolmonen, Veli-Matti and Vitucci, Enrico M. and Vainikainen, Pertti},
  journal={IEEE Trans. Antennas Propag.},
  title={Analysis and Modeling on co- and Cross-Polarized Urban Radio Propagation for Dual-Polarized MIMO Wireless Systems},
  year={2011},
  volume={59},
  number={11},
  pages={4247-4256},
  doi={10.1109/TAP.2011.2164226}}

@ARTICLE{ref22,
  author={Kouyoumjian, R.G. and Pathak, P.H.},
  journal={Proc. IEEE},
  title={A uniform geometrical theory of diffraction for an edge in a perfectly conducting surface},
  year={1974},
  volume={62},
  number={11},
  pages={1448-1461},
  doi={10.1109/PROC.1974.9651}}

@ARTICLE{ref23,
  author={Luebbers, R.},
  journal={IEEE Trans. Antennas Propag.},
  title={Finite conductivity uniform {GTD} versus knife edge diffraction in prediction of propagation path loss},
  year={1984},
  volume={32},
  number={1},
  pages={70-76},
  doi={10.1109/TAP.1984.1143189}}

@ARTICLE{ref24,
  author={Koivumäki, Pasi and Karttunen, Aki and Haneda, Katsuyuki},
  journal={IEEE Trans. Antennas Propag.}, 
  title={Ray-Optics Simulations of Outdoor-to-Indoor Multipath Channels at 4 and 14 {GHz}}, 
  year={2023},
  volume={71},
  number={7},
  pages={6046-6059},
  doi={10.1109/TAP.2023.3276502}}

@article{ref25,
  title={Recommendation {ITU-R P}. 2040-1: Effects of building materials and structures on radio wave propagation above about 100 {MHz}},
  author={Uyrus, A and Turan, B},
  journal={Recomm. ITU-R P. 2040},
  volume={1},
  year={2015}
}

@ARTICLE{ref26,
  author={Li, Zhuoyin and He, Ruisi and Yang, Mi and Qi, Ziyi and Zhang, Zhong and Ai, Bo and Zhang, Haoxiang and Han, Jiahui and Fan, Jianhua},
  journal={IEEE Antennas Wirel. Propag. Lett.}, 
  title={Ray-Tracing Calibration Based on Improved Particle Swarm Optimization in {5.9 GHz} Outdoor Scenario}, 
  year={2025},
  volume={24},
  number={8},
  pages={2592-2596},
  doi={10.1109/LAWP.2025.3569713}}

@ARTICLE{ref27,
  author={Yun, Zhengqing and Iskander, Magdy F.},
  journal={IEEE Access}, 
  title={Ray Tracing for Radio Propagation Modeling: Principles and Applications}, 
  year={2015},
  volume={3},
  number={},
  pages={1089-1100},
  doi={10.1109/ACCESS.2015.2453991}}

@ARTICLE{ref28,
  author={Guo, Yulan and Wang, Hanyun and Hu, Qingyong and Liu, Hao and Liu, Li and Bennamoun, Mohammed},
  journal={IEEE Trans. Pattern Anal. Mach. Intell.	}, 
  title={Deep Learning for {3D} Point Clouds: A Survey}, 
  year={2021},
  volume={43},
  number={12},
  pages={4338-4364},
  doi={10.1109/TPAMI.2020.3005434}}

@INPROCEEDINGS{ref29,
  author={Niu, Han and Dupleich, Diego and Völker-Schöneberg, Yanneck and Ebert, Alexander and Müller, Robert and Eichinger, Joseph and Artemenko, Alexander and Galdo, Giovanni Del and Thomä, Reiner S.},
  booktitle={Proc. IEEE 95th Veh. Technol. Conf. (VTC-Spring)}, 
  title={From {3D} Point Cloud Data to Ray-tracing Multi-band Simulations in Industrial Scenario}, 
  year={2022},
  volume={},
  number={},
  pages={1-5},
  doi={10.1109/VTC2022-Spring54318.2022.9861002}}

@INPROCEEDINGS{ref30,
  author={Yun, Zhengqing and Iskander, Magdy F.},
  booktitle={Proc. IEEE Int. Symp. Antennas Propag. USNC-URSI Radio Sci. Meeting (USNC-URSI)}, 
  title={Simplifying Building Structures for Efficient Radio Propagation Modeling}, 
  year={2023},
  volume={},
  number={},
  pages={449-450},
  doi={10.1109/USNC-URSI52151.2023.10238157}}

@ARTICLE{ref31,
  author={Meng, Wei and Li, Juan and Xi, Yong-Ji and Guo, Li-Xin and Li, Zi-Hao and Wen, Shun-Kang},
  journal={IEEE Trans. Antennas Propag.}, 
  title={An Improved Shooting and Bouncing Ray Method Based on Blend-Tree for {EM} Scattering of Multiple Moving Targets and Echo Analysis}, 
  year={2024},
  volume={72},
  number={3},
  pages={2723-2737},
  doi={10.1109/TAP.2024.3352238}}

@ARTICLE{ref32,
  author={Jürvelüinen, J. and Haneda, K.},
  journal={Radio Sci.}, 
  title={Sixty gigahertz indoor radio wave propagation prediction method based on full scattering model}, 
  year={2014},
  volume={49},
  number={4},
  pages={293-305},
  doi={10.1002/2013RS005290}}

@ARTICLE{ref33,
  author={Pang, Mingjie and Wang, Han and Lin, Kaiwei and Lin, Hai},
  journal={IEEE Antennas Wirel. Propag. Lett.}, 
  title={A {GPU}-Based Radio Wave Propagation Prediction With Progressive Processing on Point Cloud}, 
  year={2021},
  volume={20},
  number={6},
  pages={1078-1082},
  doi={10.1109/LAWP.2021.3072242}}

@misc{ref34,
      title={Photon Splatting: A Physics-Guided Neural Surrogate for Real-Time Wireless Channel Prediction}, 
      author={Ge Cao and Gabriele Gradoni and Zhen Peng},
      year={2025},
      eprint={2507.04595},
      archivePrefix={arXiv},
      primaryClass={cs.LG},
      url={https://arxiv.org/abs/2507.04595}, 
}

@ARTICLE{ref35,
  author={Zhang, Zhengyu and He, Ruisi and Yang, Mi and Zhang, Xuejian and Qi, Ziyi and Mi, Hang and Sun, Guiqi and Yang, Jingya and Ai, Bo},
  journal={Chin. J. Electron.}, 
  title={Non-Stationarity Characteristics in Dynamic Vehicular ISAC Channels at 28 GHz}, 
  year={2025},
  volume={34},
  number={1},
  pages={73-81},
  doi={10.23919/cje.2024.00.003}}

@ARTICLE{ref36,
  author={Huang, Chen and He, Ruisi and Ai, Bo and Molisch, Andreas F. and Lau, Buon Kiong and Haneda, Katsuyuki and Liu, Bo and Wang, Cheng-Xiang and Yang, Mi and Oestges, Claude and Zhong, Zhangdui},
  journal={IEEE Trans. Antennas Propag.}, 
  title={Artificial Intelligence Enabled Radio Propagation for Communications—Part {II}: Scenario Identification and Channel Modeling}, 
  year={2022},
  volume={70},
  number={6},
  pages={3955-3969},
  doi={10.1109/TAP.2022.3149665}}

@ARTICLE{ref37,
  author={He, Ruisi and Cicco, Nicola D. and Ai, Bo and Yang, Mi and Miao, Yang and Boban, Mate},
  journal={IEEE Wireless Commun.}, 
  title={{COST CA20120 INTERACT} Framework of Artificial Intelligence-Based Channel Modeling}, 
  year={2025},
  volume={32},
  number={4},
  pages={200-207},
  doi={10.1109/MWC.010.2400253}}

@ARTICLE{ref38,
  author={Zheng, Qingbi and He, Ruisi and Ai, Bo and Huang, Chen and Chen, Wei and Zhong, Zhangdui and Zhang, Haoxiang},
  journal={IEEE Wireless Commun. Lett.}, 
  title={Channel Non-Line-of-Sight Identification Based on Convolutional Neural Networks}, 
  year={2020},
  volume={9},
  number={9},
  pages={1500-1504},
  doi={10.1109/LWC.2020.2994945}}

@ARTICLE{ref39,
  author={He, Ruisi and Li, Qingyong and Ai, Bo and Geng, Yang Li-Ao and Molisch, Andreas F. and Kristem, Vinod and Zhong, Zhangdui and Yu, Jian},
  journal={IEEE Trans. Wireless Commun.}, 
  title={A Kernel-Power-Density-Based Algorithm for Channel Multipath Components Clustering}, 
  year={2017},
  volume={16},
  number={11},
  pages={7138-7151},
  doi={10.1109/TWC.2017.2740206}}

@ARTICLE{ref40,
  author={Zhang, Zhong and He, Ruisi and Yang, Mi and Qi, Ziyi and Li, Zhuoyin and Ai, Bo and Zhang, Haoxiang and Han, Jiahui},
  journal={IEEE Internet Things J.}, 
  title={Impact of Point Cloud Reconstruction Detail on {mmWave} Ray-Tracing in Indoor Environments}, 
  year={2025},
  volume={12},
  number={24},
  pages={54859-54872},
  doi={10.1109/JIOT.2025.3620998}}

@ARTICLE{ref41,
  author={Zhang, Zhengyu and Varshney, Neeraj and Senic, Jelena and Caromi, Raied and Berweger, Samuel and Gentile, Camillo and Vitucci, Enrico M. and He, Ruisi and Degli-Esposti, Vittorio},
  journal={IEEE Trans. Antennas Propag.}, 
  title={Deep Learning-based Human Gesture Channel Modeling for Integrated Sensing and Communication Scenarios}, 
  year={2025},
  volume={},
  number={},
  pages={1-1},
  doi={10.1109/TAP.2025.3648130}}

@ARTICLE{ref42,
  author={Zhang, Yuxin and He, Ruisi and Ai, Bo and Yang, Mi and Chen, Ruifeng and Wang, Chenlong and Zhang, Zhengyu and Zhong, Zhangdui},
  journal={China Commun.}, 
  title={Generative adversarial networks based digital twin channel modeling for intelligent communication networks}, 
  year={2023},
  volume={20},
  number={8},
  pages={32-43},
  doi={10.23919/JCC.fa.2023-0206.202308}}
\end{document}